\documentclass[12pt]{wlscirep}
\usepackage[utf8]{inputenc}
\usepackage[T1]{fontenc}
\usepackage{bm}
\usepackage{marvosym}
\usepackage{setspace}
\title{Observation of flat-band and band transition in the synthetic space}

\author[1,*]{Guangzhen Li}
\author[1,*]{Luojia Wang}
\author[1]{Rui Ye}
\author[1]{Shijie Liu}
\author[1,2]{Yuanlin Zheng}
\author[1\Letter]{Luqi Yuan}
\author[1,2,3,4\Letter]{Xianfeng Chen}

\affil[1]{State Key Laboratory of Advanced Optical Communication Systems and Networks, School of Physics and Astronomy, Shanghai Jiao Tong University, Shanghai 200240, China}
\affil[2]{Shanghai Research Center for Quantum Sciences, Shanghai 201315, China}
\affil[3]{Jinan Institute of Quantum Technology, Jinan 250101, China}
\affil[4]{Collaborative Innovation Center of Light Manipulation and Applications, Shandong Normal University, Jinan 250358, China}

\affil[*]{These authors contribute equally to this work.}
\affil[\Letter]{yuanluqi@sjtu.edu.cn; xfchen@sjtu.edu.cn}

\begin{abstract}
Constructions of synthetic lattices in photonics attract growingly attentions for  exploring interesting physics beyond the geometric dimensionality, among which modulated ring resonator system has been  proved as a powerful platform to create different kinds of connectivities between resonant modes along the synthetic frequency dimension with many theoretical proposals. Various experimental realizations are investigated  in a single  ring resonator, while  building beyond simple synthetic lattices in multiple rings with different types remains lacking, which desires to be accomplished as an important step further. Here, we implement the experimental demonstration of generating the one-dimensional  Lieb lattice along the frequency axis of light, realized in two coupled ring resonators while the larger ring undergoing dynamic modulation. Such synthetic photonic structure naturally exhibits the physics of flat band. We show that the time-resolved band structure read out from the drop-port output of the  excited ring is the intensity projection of the band structure onto specific resonant mode in the synthetic momentum space, where  gapless flat band, mode localization effect, and flat to non-flat band transition are  observed in experiments and verified by simulations. Our work  gives a direct evidence for the  constructing synthetic Lieb lattice with two rings, which hence makes a  solid step towards experimentally constructing more complicated lattices in multiple rings  associated with synthetic frequency dimension.
\end{abstract}

\begin{document}

\flushbottom
\maketitle

\thispagestyle{empty}

\section*{Main text}

Synthetic dimension in photonics attracts broad interest and experiences important experimental achievements in recent years \cite{Yuan:18,ozawa2019topological,Lustig:21,yuan2021tutorial}, which shows the great capability for studying fundamental physical phenomena with exotic artificial connectivities \cite{Schwartz:13,qin2018spectrum,yuan2018prb,lustig2019photonic,wang2020multidimensional}, manipulating light in  various ways \cite{regensburger2011photon,regensburger2012parity,luo2015quantum,bell2017spectral,ozawa2017synthetic,peterson2019strong,yuan2019photonic,yuan2021prr}, and pointing towards exploring higher-dimensional physics beyond three dimensions  \cite{PhysRevA.87.013814,PhysRevB.98.125431,zilberberg2018photonic}. Among these experimental achievements, different degrees of freedom of light, including arrival times of pulses \cite{regensburger2011photon,regensburger2012parity}, frequencies \cite{yuan2016photonic,ozawa2016synthetic,dutt2019experimental}, and modal dimensions \cite{lustig2019photonic}, have been used to construct the synthetic dimension. Hence, a variety of novel physics have been demonstrated in synthetic dimensions, such as the photonic topological insulator \cite{lustig2019photonic}, the Hall ladder with the effective magnetic flux \cite{dutt2020a}, the trajectory of dynamic band structures \cite{Lieabe4335}, and the topological funneling with non-Hermitian physics \cite{weidemann2020topological}, whose physical models are hard to be built in structures with only spatial dimensions.

Besides different photonic platforms for constructing synthetic dimensions, dynamically modulated ring resonator system  has manifested as a powerful platform where   resonant modes  with equally-spaced frequencies are coupled by the external modulation and then synthetic frequency dimension is created \cite{yuan2016photonic,ozawa2016synthetic}. The modulation applied by external voltages provides the unique advantages of breaking the constrain of fixed geometric structures after fabrication and thus provides an important possibility of achieving complicated functionalities  with great experimental flexibility and reconfigurability \cite{yuan2021tutorial}. Up to now,  experimental implementations  in ring resonator systems including fiber loops or on-chip  microrings  have demonstrated  the creation of the synthetic frequency dimension in a single ring  resonator \cite{dutt2019experimental,dutt2020a,Hu:20,chen2021real,Lieabe4335,wang2021generating,balvcytis2021synthetic},  where versatile physical phenomena have been  shown, such as measuring band structures \cite{dutt2019experimental,Lieabe4335}, observing spectral Bloch oscillations \cite{chen2021real}, and generating arbitrary topological windings \cite{wang2021generating}. On the other hand, many theoretical proposals have been explored where formations of photonic lattice   with different types of rings in higher dimensions can intrigue studies of rich physics, such as simulating two-dimensional  Haldane model \cite{yuan2018prb}, three-dimensional topological insulator \cite{Lineaat2774}, and four-dimensional quantum Hall effect \cite{ozawa2016synthetic} in multiple rings. However, to realize these theoretical proposals, it requires constructing a synthetic space including the frequency dimension in two or more rings \cite{wangkai2021nature} with perhaps different types, which are still missing in experiments, due to the fact that resonant modes circulated in each ring  at different type need fully precise synchronization in order to stably connect same synthetic frequency sites in different rings.  In other words, the experimental feasibility for constructing synthetic space in multiple rings with different types  are questionable so far. Therefore, as a crucial step towards exploring complex lattice structures in the synthetic space with multiple rings, one desires the demonstration of creating the synthetic space in two coupled rings with different types  in the experiment.

In this work, we prove the capability for coupling two rings with different types where one ring undergoes the dynamic modulation, and observe flat-band physics in a synthetic space including the frequency dimension. Such configuration supports a one-dimensional (1D) photonic Lieb lattice but associated with synthetic frequency dimension. One intrinsic physics of such lattice is the naturally existence of the flat (dispersionless) band, which has found important applications with the localization effect \cite{mukherjee2015observation,baboux2016bosonic,leykam2018artificial,PhysRevLett.121.263902,PhysRevLett.124.183901}.
In our experiments,  the time-resolved energy bands from the drop-port output of the excited ring are obtained, corresponding to the projection of  the band structures of the 1D Lieb lattice, which, however, is on the synthetic dimension. Moreover,  by exciting the resonant modes through the selected input port of one ring and recording the output transmission  from the same ring, we observe the effective localization of the resonant modes near the flat band. Such flat band in the synthetic space can further be modified by adding the long-range couplings in the modulation, which leads to the transition from the flat to non-flat bands. We provide theoretical analysis, which gives excellent agreements with experiments.  Our work explores the flat-band physics in a synthetic Lieb lattice along the frequency dimension by successfully synchronizing two rings with different types, which exhibits a crucial step towards the potential for experimentally constructing more complicated synthetic lattices in multiple rings.

\begin{figure}[htb]
\centering
\includegraphics[width=9 cm]{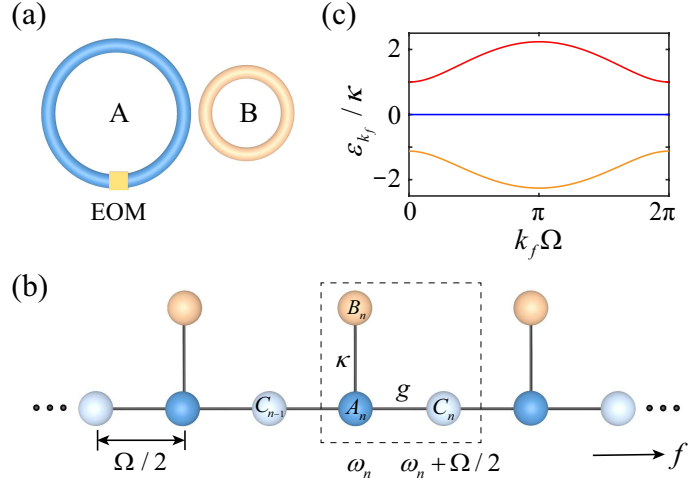}
\captionsetup{font={small,stretch=1.5}, justification=raggedright}
\caption{\label{fig1} \textbf{Configuration of a 1D synthetic photonic Lieb lattice.} (a) Two coupled ring resonators, where the FSR of ring A is half of the FSR of ring B, i.e., $2\Omega_{\mathrm{A}}=\Omega_{\mathrm{B}}\equiv\Omega$.   Ring A undergoes dynamic modulation by placing an EOM with the modulation frequency $\Omega_{\mathrm{M}}=\Omega/2$. (b) The system in (a) can be mapped into a 1D photonic Lieb  lattice along the synthetic frequency dimension ($f$), with $A_n$, $B_n$, and $C_n$ indicating three types of lattice sites. (c) The corresponding  band structures of the synthetic Lieb lattice in (b) with $g=\kappa$ and $\phi=-0.5\pi$.}
\end{figure}

We start with  illustrating the  model including two types of ring resonators  composed by the waveguide or fiber loop fabricated with same materials at different lengths, labelled as A and B  as shown in Fig.~\ref{fig1}(a). In the absence of the group velocity dispersion, the ring resonator supports a set of modes with equally spaced frequencies. If we set the central resonant frequency at $\omega_0$, the $n^{\mathrm{th}}$  mode in the ring A(B) has the  frequency $\omega_{\mathrm{A(B)},n}=\omega_0+n\Omega_{\mathrm{A(B)}}$, where $\Omega_{\mathrm{A(B)}}=2\pi v_g/L_{\mathrm{A(B)}}$ is the free spectral range (FSR) of ring A(B), and $v_g$ is the group velocity. We consider the length of ring A ($L_{\mathrm{A}}$)  twice long as the length of  ring B ($L_{\mathrm{B}}$), i.e., $L_{\mathrm{A}}=2L_{\mathrm{B}}$, which gives $2\Omega_{\mathrm{A}}=\Omega_{\mathrm{B}}\equiv\Omega$. There is an electro-optic modulator (EOM) placed inside the ring A with the modulation frequency $\Omega_{\mathrm{M}}=\Omega/2$, the modulation strength $g$, and the modulation phase $\phi$, which provides the connectivity between adjacent resonant modes in ring A, while there is no modulator in ring B so  resonant modes in ring B remain un-connected. Two resonant modes in two rings at the same frequency  can be coupled  through evanescent waves or fiber coupler with the coupling strength $\kappa$ [see Fig.~\ref{fig1}(b)]. Therefore, there exist three types of modes in the system, defined as $A_n$, $B_n$, and $C_n$, where  $A_n$ and $C_n$ are the resonant modes at frequencies $\omega_n$ and  $\omega_n+\Omega/2$  in ring A,  and $B_n$ is the resonant mode at frequency $\omega_{n}$ in ring B with $\omega_{n}=\omega_0+n\Omega$.   In particular, modes $A_n$ and $B_n$ are coupled for the same $n$, while $A_n$ is coupled to $C_{n-1}$ and $C_n$ through the modulation under the lowest-order approximation, resulting the synthetic lattice shown in Fig.~\ref{fig1}(b).

The corresponding tight-binding Hamiltonian of the system is \cite{yu2020isolated}
\begin{equation}\label{eq1}
\begin{split}
H & =H_0+H_1=\sum_{n}\Big[\omega_n\Big(a^{\dag}_na_n+b^{\dag}_nb_n\Big)+(\omega_n+\Omega/2)c^{\dag}_nc_n\Big]\\
& +\kappa\sum_{n} \Big(a^{\dag}_nb_n+h.c.\Big)+2g\cos(\Omega t/2+\phi)\sum_{n}\Big(a^{\dag}_nc_n+a^{\dag}_nc_{n-1}+h.c.\Big),
\end{split}
\end{equation}
where $a^{\dag}_n$, $b^{\dag}_n$, and $c^{\dag}_n$ ($a_n$, $b_n$, and $c_n$) are the  creation (annihilation) operators for the modes $A_n$, $B_n$ and $C_n$, respectively.
$H_0$ denotes the first summation terms on the right of Eq.~(\ref{eq1}), referring to the unperturbed part for all resonant modes, while $H_1$ gives the interaction terms.
Equation~(\ref{eq1}) can be simplified into the interaction picture by considering $e^{iH_0t}H_1e^{-iH_0t}$, and taking  the rotating-wave approximation \cite{scully_zubairy_1997}, which results in
\begin{equation}\label{eq2}
H_c=\kappa\sum_{n}a^{\dag}_nb_n+ g\sum_{n}\Big(a^{\dag}_nc_ne^{i\phi}+a^{\dag}_nc_{n-1}e^{-i\phi}\Big)+h.c..
\end{equation}
Equation~(\ref{eq2}) describes the Hamiltonian of a  synthetic  lattice structure, which is analog to the 1D spatial Lieb lattice \cite{hyrkas2013many,PhysRevA.96.043803,huda2020designer}, but it is along the frequency axis of light.

To understand the underlying physics of the Hamiltonian described in Eq.~(\ref{eq2}), we can re-write Eq.~(\ref{eq2}) into the $k_f$ space
\begin{equation}\label{eq3}
H_k=\kappa\sum_{k_f}\Big(a^{\dag}_{k_f}b_{k_f}+b^{\dag}_{k_f}a_{k_f}\Big)+2g\sum_{k_f}\Big(a^{\dag}_{k_f}c_{k_f}+c^{\dag}_{k_f}a_{k_f}\Big)\cos(k_f\Omega/2+\phi),
\end{equation}
where $k_f$ is the wave vector reciprocal to the frequency dimension acting as a time variable \cite{yuan2021tutorial}. The corresponding  photonic band structure  of the system is then given by
\begin{equation}\label{eq8}
  \varepsilon_{k_f,0}=0,~\varepsilon_{k_f,\pm}=\pm\sqrt{[2g\cos(k_f\Omega/2+\phi)]^2+\kappa^2},
\end{equation}
where $\varepsilon_{k_f,j} (j=0,\pm)$   are the  eigenvalues from Eq.~(\ref{eq3}), corresponding to three bands    plotted in Fig.~\ref{fig1}(c) within the first Brillouin zone with $k_f\in[0,2\pi/\Omega]$. One can see  a flat band $\varepsilon_{k_f,0}$ in the middle gapped from the upper and lower dispersive bands $\varepsilon_{k_f,\pm}$, which indicates that light can be efficiently localized in the flat band without scattering \cite{mukherjee2015observation,baboux2016bosonic,leykam2018artificial,PhysRevLett.121.263902,PhysRevLett.124.183901}.
Let $\psi_{k_f,j}=(\psi^A_{k_f,j},\psi^B_{k_f,j},\psi^C_{k_f,j})^{\mathrm{T}}$ be the   eigenstates corresponding to $\varepsilon_{k_f,j}$, with $\psi^A_{k_f,j}$, $\psi^B_{k_f,j}$ and $\psi^C_{k_f,j}$ being the projection of the eigenstates  on the three modes ($A_k$, $B_k$, and $C_k$) in the $k_f$ space, and then we have
\begin{equation}\label{eq9}
\begin{split}
\psi_{k_f,0} & = \big(0,-G,\kappa\big)^{\mathrm{T}}/\sqrt{G^2+\kappa^2},\\
  \psi_{k_f,\pm} & =\big(\pm\sqrt{G^2+\kappa^2},\kappa,G\big)^{\mathrm{T}}/\sqrt{2(\kappa^2+G^2)},
\end{split}
\end{equation}
%\begin{eqnarray}
%% \nonumber to remove numbering (before each equation)
%(a_{k_f,j}) &=& (0,1,-1)/\sqrt{2}\label{eq9}, \\
% (b_{k_f,j}) &=& (-\sqrt{2}G,\kappa,\kappa)/\sqrt{2(\kappa^2+G^2)}\label{eq11},\\
% (c_{k_f,j}) &=& (\sqrt{2}\kappa,G,G)/\sqrt{2(\kappa^2+G^2)}\label{eq10},
%\end{eqnarray}
with $G=2g\cos(k_f\Omega/2+\phi)$. One notes that the flat band ($j=0$) has no projection onto the mode $A_k$ due to $\psi^A_{k_f,0}=0$, while the two dispersive bands ($j=\pm$) are asymmetrically projected onto the mode $A_k$ but symmetrically projected onto modes $B_k$ and $C_k$.

\begin{figure*}[htb]
\center
\includegraphics[width=14 cm]{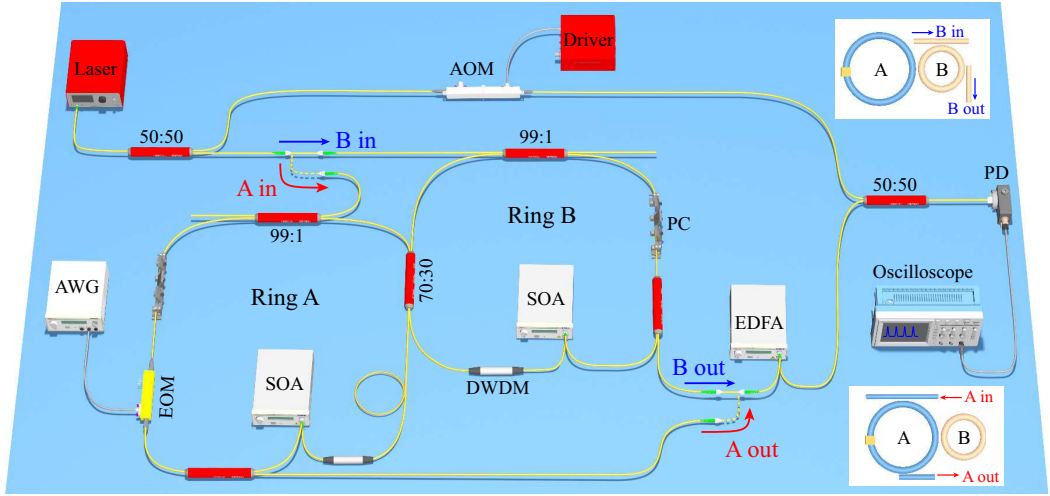}
\captionsetup{font={small,stretch=1.5}, justification=raggedright}
\caption{\label{fig2} \textbf{Experimental setup. } EOM: electro-optic phase modulator. AOM: acousto-optic modulation. SOA: semiconductor optical amplifier. AWG: arbitrary waveform generator. EDFA: erbium-doped optical fiber amplifier. PC: polarization controller. DWDM: dense wavelength division multiplexing. PD: photodiode. Inserted:   sketches of B $in$$\rightarrow$B $out$ and A $in$$\rightarrow$A $out$.}
\end{figure*}

To implement the idea of the 1D synthetic photonic Lieb lattice described in Eq.~(\ref{eq2})  for the potential experimental demonstration, we continue with considering a realistic model of two ring resonators coupled with input and output waveguides.  In the following, we consider two excitation cases by selectively choosing the input/output ports, which are referred as  B $in$$\rightarrow$B $out$  and A $in$$\rightarrow$A $out$  as sketched in the inserted  figures in Fig.~\ref{fig2}. First, we inject the field into the system through the input port of ring B and measure drop-port output of ring B as well (B $in$$\rightarrow$B $out$), in which way  only frequency mode $B_n$ in ring B is directly  excited. We consider the photon state being $|\psi\rangle=\sum_n\big[v_{a,n}(t)a_n^{\dagger}+v_{b,n}(t)b_n^{\dagger}+v_{c,n}(t)c_n^{\dagger}\big]|0\rangle$, where $v_{a,n}$, $v_{b,n}$, and  $v_{c,n}$ are the amplitudes of the photon states of the frequency modes $A_n$, $B_n$, and $C_n$, respectively. By  applying the Schr$\ddot{\mathrm{o}}$dinger equation $i|\dot{\psi}\rangle=H_c|\psi\rangle$ with $H_c$ defined in Eq.~(\ref{eq2}), we obtain the input/output coupled amplitude equations for the $n^{\mathrm{th}}$ modes \cite{yuan2021tutorial}
\begin{equation}\label{eq4}
  \begin{split}
     \dot{v}_{a,n} & =-i\kappa v_{b,n}-ig\big(v_{c,n}e^{i\phi}+v_{c,n-1}e^{-i\phi}\big)-\gamma v_{a,n}, \\
     \dot{v}_{c,n} & =-ig\big(v_{a,n}e^{-i\phi}+v_{a,n+1}e^{i\phi}\big)-\gamma v_{c,n}, \\
     \dot{v}_{b,n} & =-i\kappa v_{a,n}-\gamma v_{b,n}+i\sqrt{\gamma_{\mathrm{B}}}S_{\mathrm{in}}^{\mathrm{B}}e^{in\Omega t-i\Delta\omega t},
  \end{split}
\end{equation}
where $S_{\mathrm{in}}^{\mathrm{B}}$ is the input laser source, $\Delta\omega$ is the frequency detuning between the input field frequency and the reference frequency $\omega_0$, $\gamma$ is the total loss, and $\gamma_{\mathrm{B}}$ is the  coupling strength between ring B and waveguides. The  amplitude of the output field at the drop-port of ring B is given by $S_{\mathrm{out}}^{\mathrm{B}}=-i\sqrt{\gamma_{\mathrm{B}}}\sum_{n}v_{b,n}e^{-i\omega_nt}$. The normalized drop-port transmission $T_{\mathrm{out}}^{\mathrm{B}}=|S_{\mathrm{out}}^{\mathrm{B}}/S_{\mathrm{in}}^{\mathrm{B}}|^2$ can be expressed in the $k_f$ space as [see supplementary materials]
\begin{equation}\label{eq5}
T_{\mathrm{out}}^{\mathrm{B}}(t=k_f;\Delta\omega)= \gamma_{\mathrm{B}}^2\frac{|\psi^B_{k_f,j}|^4}{(\Delta\omega-\varepsilon_{k_f,j})^2+\gamma^2},
\end{equation}
with $\varepsilon_{k_f,j}$ and $\psi^B_{k_f,j}$ being determined by Eqs.~(\ref{eq8})-(\ref{eq9}).
Previous works have demonstrated that the photonic band structure can be measured by time-resolved transmission spectroscopy, where the drop-port output transmission signal is obtained  by scanning the frequency of the input laser linearly with time \cite{dutt2019experimental,Lieabe4335}.
Therefore, Eq.~(\ref{eq5}) indicates that the band structures read out from the drop-port output of ring B exhibit the projection of  the band structure on the mode $B_k$ in $k_f$ space.

On the other hand, for the case of A $in$$\rightarrow$A $out$ by changing the input/output port to ring A, similar input/output coupled amplitude equations can be obtained. The  amplitude  of the output field at the drop port in the ring A is  described by $S_{\mathrm{out}}^{\mathrm{A}}=-i\sqrt{\gamma_{\mathrm{A}}}\sum_{n}\big[v_{a,n}e^{-i\omega_nt}+v_{c,n}e^{-i(\omega_n+\Omega/2)t}\big]$, with $\gamma_{\mathrm{A}}$ being the waveguide-resonator  coupling strength of ring A.
The corresponding normalized drop-port transmissions  in the $k_f$ space are [see supplementary materials]
 \begin{equation}\label{eq6}
  T_{\mathrm{out}}^{\mathrm{A}}(t=k_f;\Delta\omega)=\gamma_{\mathrm{A}}^2\frac{|\psi^A_{k_f,j}|^2~|\psi^A_{k_f,j}+\psi^C_{k_f,j}|^2}{(\Delta\omega-\varepsilon_{k_f,j})^2+\gamma^2},
 \end{equation}
 \begin{equation}\label{eq7}
T_{\mathrm{out}}^{\mathrm{A}}(t=k_f;\Delta\omega+\Omega/2)=\gamma_{\mathrm{A}}^2\frac{|\psi^C_{k_f,j}|^2~|\psi^A_{k_f,j}+\psi^C_{k_f,j}|^2}{(\Delta\omega-\varepsilon_{k_f,j})^2+\gamma^2},
\end{equation}
where Eq.~(\ref{eq6}) and Eq.~(\ref{eq7}) refer to the situation of an input field near resonant with the reference frequency $\omega_0$ and $\omega_0+\Omega/2$, respectively. This means that the band structure resolved from the drop-port transmission through the ring A is the projection of the band structure on the modes $A_k$ and $C_k$ separated by $\Omega/2$ along the frequency dimension.

Different from previous experiments conducted in a single ring, the construction of the synthetic Lieb lattice requires two coupled rings, which faces the challenge of fully synchronizing resonant modes in two rings while one ring is undergoing dynamic modulation. In experiments, we use two fiber ring resonators coupled together through a $2\times2$ fiber coupler with coupler ratio 70:30 as shown in Fig.~\ref{fig2} (see
Methods). The two rings are excited separately by selectively choosing  ring A or ring B as the input port of the laser source (A $in$ or B $in$), while the transmission is recorded from the corresponding drop port (A $out$ or B $out$). After calibration, the lengths of the two rings are $L_{\mathrm{A}}=20.4$ m and $L_{\mathrm{B}}=10.2$ m, corresponding to  $\Omega_{\mathrm{A}}=2\pi\cdot10$ MHz and $\Omega_{\mathrm{B}}=2\pi\cdot20$ MHz.  To form the synthetic Lieb lattice described in Fig.~\ref{fig1}(b), we drive the EOM in ring A by a sinusoidal radio frequency (RF) signal in the form of $V_{\mathrm{M}}\cos(\Omega_{\mathrm{M}}t+\phi) $ with $\Omega_{\mathrm{M}}=2\pi\cdot10$ MHz, and $\phi=-0.5\pi$.

\begin{figure}[htb]
\center
\includegraphics[width=12 cm]{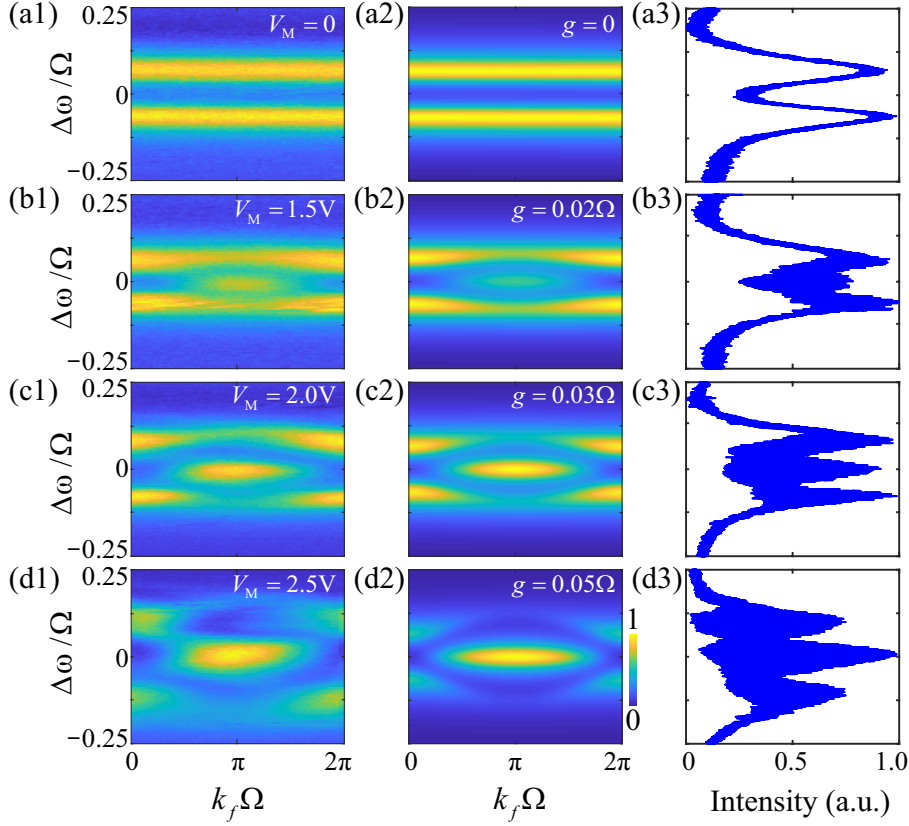}
\captionsetup{font={small,stretch=1.5}, justification=raggedright}
\caption{\label{fig3}  \textbf{Band structure measurements for the case of B $\emph{\textbf{in}}$$\rightarrow$B $\emph{\textbf{out}}$.} (a1)-(d1) Experimentally observed band structures  with different modulation amplitudes $V_{\mathrm{M}}$. (a2)-(d2) Simulation results of the projected output intensity distribution of  the band structure on mode $B_k$,  based on Eqs.~(\ref{eq8})-(\ref{eq9}) and (\ref{eq5}), where  $g$ takes different values with  fixed  $\kappa=0.06\Omega$  and $\phi=-0.5\pi$. (a3)-(d3) Measured transmission spectra from the drop port of ring B. The vertical axis represents the frequency detuning of the input laser source normalized to $\Omega$, while the bottom horizontal  axis in (a1)-(d2) represents one roundtrip time in ring B with the period of $2\pi/\Omega$. }
\end{figure}

To   demonstrate the construction of the synthetic photonic Lieb lattice in the experiment, we perform the band structure measurements by finely sweeping the frequency of the input laser through multiple free-spectral ranges \cite{Lieabe4335}. We first inject the laser source into the input  port of ring B and  measure the output transmission spectra from the drop port of ring B (B $in$$\rightarrow$B $out$). Figure \ref{fig3}(a1)-(d1) plot the time-resolved band structures with varying modulation amplitude $V_{\mathrm{M}}$, which are obtained by breaking the measured output transmission signals  in Figs.~\ref{fig3}(a3)-(d3) into time slices with the time window equaling to one roundtrip time  of ring B ($2\pi/\Omega$), i.e., the periodicity of the synthetic Lieb lattice. We calculate the intensity projections of the band structure on the mode $B_k$ using Eq.~(\ref{eq5}) and show the results in  Figs.~\ref{fig3}(a2)-(d2).
Without modulation ($V_{\mathrm{M}}=0$), one  sees that coupled rings result in two Lorentzian resonances of the unmodulated rings which are slightly overlapped due to the  lineshape broadening from the loss of the system [see Fig.~\ref{fig3}(a3)]. It leads to two straight energy bands  with constant intensity distributions in both experiment [see Fig.~\ref{fig3}(a1)] and theory [see Fig.~\ref{fig3}(a2)].
The feature of synthetic Lieb lattice  begins to manifest once the modulation is applied as shown in  Figs.~\ref{fig3}(b1)-(d1), where one notices that there exists three bands and the intensity distributions vary with the modulation amplitude.
For a small modulation amplitude [see Fig.~\ref{fig3}(b1) with $V_{\mathrm{M}}=1.5$ V], the energy of the eigenstate mainly  focuses on the upper and lower dispersive bands, which  transfers to the middle flat band when the  modulation strength  becomes larger as shown in Fig.~\ref{fig3}(d1) with $V_{\mathrm{M}}=2.5$ V. The theoretical plots exhibit excellent agreement with experimental measurements, which clearly shows that the energy of the eigenstate flows from dispersive bands to the flat band when increasing $g$ [see Fig.~\ref{fig3}(b2)-(d2)]. Moreover, the  intensity distributions on the two dispersive bands have symmetric pattern within the first Brillouin zone of $k_f\in[0,2\pi/\Omega]$, which is consistent to  the analytical solutions in  Eq.~(\ref{eq9}).

\begin{figure}[htb]
\center
\includegraphics[width=12 cm]{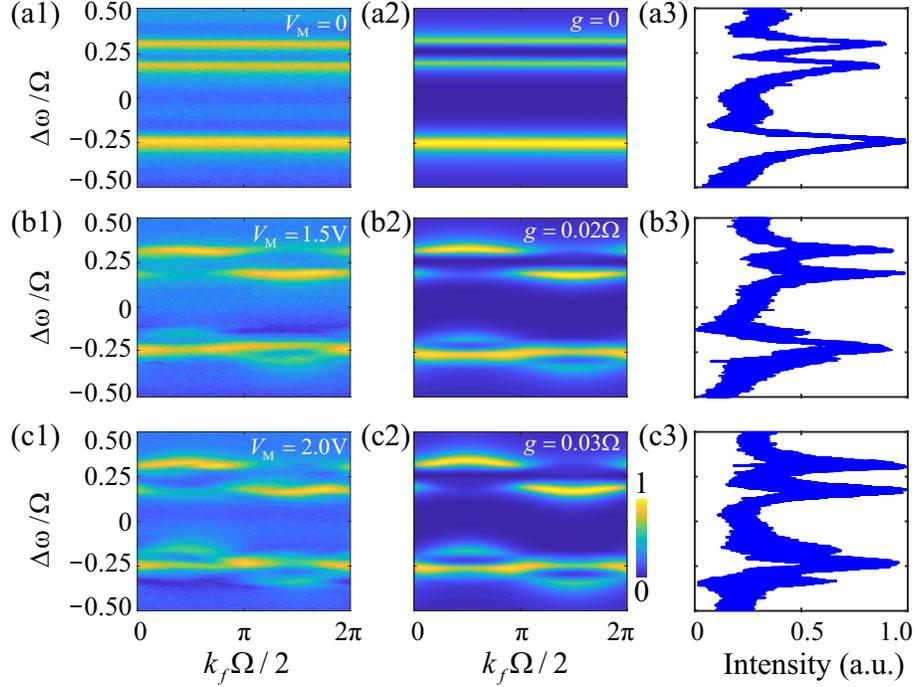}
\captionsetup{font={small,stretch=1.5}, justification=raggedright}
\caption{\label{fig4} \textbf{Band structure measurements for the case of A $\emph{\textbf{in}}$$\rightarrow$A $\emph{\textbf{out}}$.} (a1)-(c1) Experimentally observed band structures varied with  $V_{\mathrm{M}}$. (a2)-(c2) Simulation results of the projected  intensity distribution of  the  band structure on modes $A_k$ and $C_k$, based on Eqs.~(\ref{eq8})-(\ref{eq9}) and (\ref{eq6})-(\ref{eq7}), with $\kappa=0.06\Omega$   and $\phi=-0.5\pi$. (a3)-(c3) Transmission spectra measured from the drop port of ring A.  The bottom horizontal axis in (a1)-(c2) represents one roundtrip time in ring A with the period of $4\pi/\Omega$.}
\end{figure}

We then consider the case of A $in$$\rightarrow$A $out$ by switching the  input and output fiber to ring A. The output transmissions of modes $A_k$ and $C_k$ separated by $\Omega/2$ are measured simultaneously as shown in Figs.~\ref{fig4}(a3)-(c3). Since in experiments, the time window to break the  measured output transmission signals  of ring A  equals to  one roundtrip time  of ring A ($4\pi/\Omega$), we measure a combination of intensity projections of the band structure on $A_k$ and $C_k$, which gives   $k_f\in[0,4\pi/\Omega]$ as plotted in Figs.~\ref{fig4}(a1)-(c1).
Theoretical results from Eqs.~(\ref{eq6})-(\ref{eq7}) are plotted in Figs.~\ref{fig4}(a2)-(c2). When there is no modulation, one sees two nearby straight bands near $\Delta\omega/\Omega=0.25$ due to the energy splitting from coupling between modes $A_n$ and $B_n$, and one single straight band near $\Delta\omega/\Omega=-0.25$ referring to the resonance of $C_n$ in both experiment and theory [see Fig.~\ref{fig4}(a1)-(a2)].
When the modulation is applied, the band structures near  $\Delta\omega/\Omega=\pm0.25$ show different features. For upper bands near  $\Delta\omega/\Omega=0.25$, one sees two dispersive bands, corresponding to the band structure in Fig.~\ref{fig1}(c) projected to modes $A_k$ [see Fig.~\ref{fig4}(b1)-(c1)], which matches well with the calculated results by the formula in Eq.~(\ref{eq6}) [see Fig.~\ref{fig4}(b2)-(c2)]. On the other hand, for lower bands near  $\Delta \omega/\Omega = - 0.25$, one can clearly see three bands, with the middle one being flat. The intensity projections of two dispersive bands on mode $C_k$ are relatively weak in both experiment and theory.
Both the  intensity distributions of the two dispersive bands on the mode  $A_k$ and $C_k$ have the asymmetric patterns within one period, which matches with  the theoretical result in  Eq.~(\ref{eq9}).
We shall emphasize that the periodicity of the signal with a time window being 4$\pi/\Omega$ can also be noticed from the term $|\psi^A_{k_f,j}+\psi^C_{k_f,j}|^2$, which is a unique characteristics from our system that signal amplitudes from modes $A_n$ and $C_n$  are mixed in the experiment.
Furthermore, the roughness of transmission spectra in both Figs.~\ref{fig3} and \ref{fig4} originates from the small display of frequency detuning range for containing multiple sinusoidal signal periods \cite{Lieabe4335}.

\begin{figure}[htb]
\center
\includegraphics[width=10 cm]{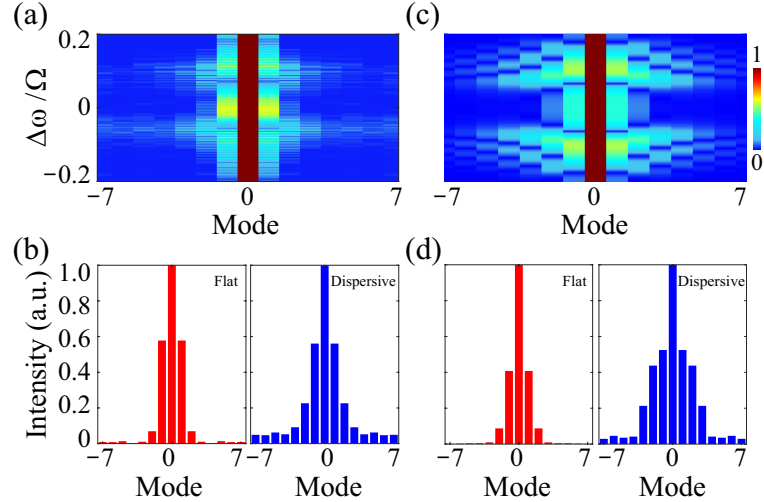}
\captionsetup{font={small,stretch=1.5}, justification=raggedright}
\caption{\label{fig5} \textbf{Mode distributions for the case of B $\emph{\textbf{in}}$$\rightarrow$B $\emph{\textbf{out}}$.} (a) Experimentally resolved resonant mode spectra as a function of frequency detuning with $V_{\mathrm{M}}=3$ V. (b) The corresponding mode distributions of two selected input frequencies in (a) located at $\Delta\omega=0$  and $\Delta\omega=0.08\Omega$, respectively. (c) Simulated resonant  mode spectra $|v_{b,n}|^2$  with $g=\kappa=0.06\Omega$, and (d) the corresponding intensity distributions of the two chosen input frequencies at $\Delta\omega=0$  and $\Delta\omega=0.08\Omega$, respectively.
The horizontal axis represents the   mode number $n$ for the frequency $\omega_n$. }
\end{figure}

Next, we   measure the frequency mode distributions for the case of B $in$$\rightarrow$B $out$ by the heterodyne detection method to probe the localization effect of  flat band  in the synthetic Lieb lattice. We connect the AOM path in Fig.~\ref{fig2} for frequency shift, and interfere it with the drop-port output of ring B by a 50:50 fiber coupler. To show  evolutions of frequency modes through out the whole band structure, we  sweep the input laser frequency  near the resonance frequency $\omega_0$, and process the drop-port output transmission through the fast Fourier transform \cite{nussbaumer1981fast}.  Figure~\ref{fig5}(a) shows the experimentally resolved mode distributions as a function of frequency detuning $\Delta\omega$, where the intensities of modes are well confined near $\Delta\omega\thicksim0$, which refers to the flat band, but spreads over the dispersive bands at $\Delta\omega\thicksim\pm0.07\Omega$. We explicitly exhibit the mode intensity distributions  for two input frequencies in Fig.~\ref{fig5}(b), which are  $\Delta\omega=0$ at the flat band and  $\Delta\omega=0.08\Omega$ at the upper dispersive band, respectively.
For the input frequency at the flat band [see the left part of Fig.~\ref{fig5}(b)], one sees that the intensities of modes $B_n$ mainly locate at the $0^{\mathrm{th}}$ and $\pm1^{\mathrm{st}}$  modes with very small portion diverging to the $\pm2^{\mathrm{nd}}$ modes. On the other hand, intensities of modes experience  spread for the input frequency located at  the dispersive band [see the right part of Fig.~\ref{fig5}(b)]. Simulations are performed by solving  Eq.~(\ref{eq5}) with sweeping the input frequency and then Fourier transforming the transmitted signal. One can see a good agreement between experimental measurement in Figs.~\ref{fig5}(a)-(b) and simulated results in  Figs.~\ref{fig5}(c)-(d).

\begin{figure}[htb]
\center
\includegraphics[width=10 cm]{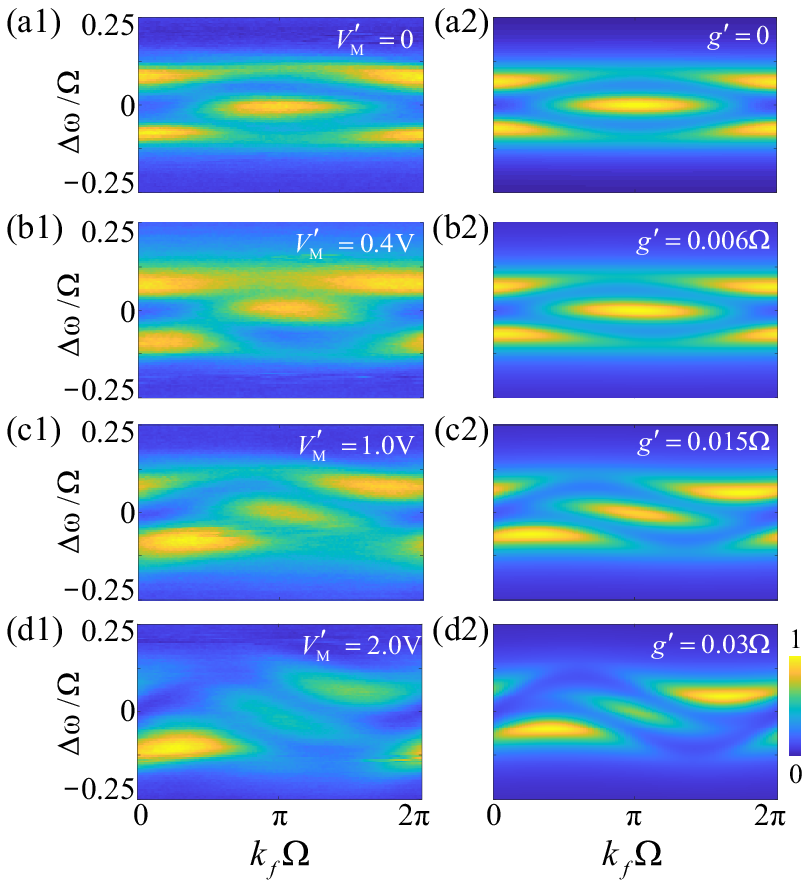}
\captionsetup{font={small,stretch=1.5}, justification=raggedright}
\caption{\label{fig6} \textbf{Observations of flat to non-flat band transition for the case of B $\emph{\textbf{in}}$$\rightarrow$B $\emph{\textbf{out}}$.} (a1)-(d1) Experimentally measured band structures with different long-range modulation amplitudes $V_{\mathrm{M}}'$ and fixed $V_{\mathrm{M}}=2$ V. (a2)-(d2) Simulation results of the projected intensity distribution of the band structure on mode $B_k$ varied with $g'$, where $g=0.03\Omega$ and $\phi=\phi'=-0.5\pi$.  }
\end{figure}

One notes that the existence of the flat band in the constructed synthetic Lieb lattice is not dependent on the coupling coefficients, i.e., $g$ and $\kappa$ for the weak modulation condition \cite{baboux2016bosonic}. Further increase of the modulation strength falls on the break of the synthetic Lieb lattice under the tight-binding limit. In this structure, one can make the band transition between the flat band and non-flat band by simply adding the higher-order modulation to introduce the long-range couplings in the frequency dimension, i.e., with an additional modulation frequency $\Omega$, which makes the modulation being $2g\cos(\Omega t/2 + \phi)+2g'\cos(\Omega t + \phi')$.  The second term in the modulation brings next-nearest-nearby couplings between two nearby resonant modes $A_n$ (or $C_n$). In experiment, we apply the EOM in ring A with the corresponding form of $V_{\mathrm{M}}\cos(\Omega_{\mathrm{M}}t+ \phi)+V_{\mathrm{M}}'\cos(2\Omega_{\mathrm{M}}t+ \phi') $ where $\Omega_{\mathrm{M}}=2\pi\cdot10$ MHz, and $\phi=\phi'=-0.5\pi$, and perform the measurements in the case of B $in$$\rightarrow$B $out$, which are shown in  Fig.~\ref{fig6}. Without higher-order modulation [see Fig.~\ref{fig6}(a1) with $V_{\mathrm{M}}'=0$], the system exhibits the feature of flat band, which is the same with Figs.~\ref{fig3}(c1)-(c2). Once the higher-order modulation term is added into EOM ($V_{\mathrm{M}}'\neq0$), the middle band gradually turns into dispersive, while the upper and lower dispersive bands start to show the nonsymmetric feature as shown in Figs.~\ref{fig6}(b1)-(d1). In addition, the gap throughout the entire $k_f$ space gets closed if $V_{\mathrm{M}}'$ becomes larger [see Fig.~\ref{fig6}(d1)]. We therefore show the transition from  flat to non-flat bands in Figs.~\ref{fig6}(a1)-(d1), which are excellently agreed with  the simulations results in Figs.~\ref{fig6}(a2)-(d2).
Such opportunity to dynamically introduce the band transition could be useful for light stopping, which has been proposed in theory \cite{Yanik2004,Sandhu:06}.

Our construction of the Lieb lattice in two coupled rings, which is more complex than the 1D lattice in the frequency dimension in a single modulated ring \cite{dutt2019experimental,dutt2020a,Hu:20,chen2021real,Lieabe4335,wang2021generating,balvcytis2021synthetic}, proves the experimental feasibility in connecting multiple rings  with different types while simultaneously  constructing synthetic frequency dimension.
Our experiments also points to important future opportunities for  experimentally implementing other complicated synthetic  lattice structures in multiple rings, such as  square lattice \cite{yuan2016photonic}, honeycomb lattice \cite{yuan2018prb}, and lattice supporting high-order topology \cite{zhang2020quadrupole,dutt2020higher}, which have been intensively studied in theory \cite{yuan2021tutorial}.  Moreover, miniaturization of modulated rings is also possible thanks to the advances of integrated on-chip platforms \cite{Hu:20,balvcytis2021synthetic}.

In summary, we have experimentally demonstrated a 1D synthetic photonic Lieb lattice along the frequency axis of light, constructed by two coupled fiber ring resonators at different types. Flat band gapped from two dispersive bands are observed under both two cases by selectively choosing the input and output ports for excitations and transmission measurements, which shows that  measured band structures  are  intensity projections of the band on the different resonant modes in $k_f$ space. We also observe  localization effect near the flat band with distinctive feature from dispersive bands, and demonstrate the flat to non-flat band transition by adding the long-range couplings in modulations, characterizing  the intrinsic physics of the  synthetic Lieb lattice. Theoretical simulations are performed and agree well with experimental results, which shows unique feature of synthetic frequency dimension in measurements  where signal from two modes are combined.
 Our work opens a door for experimentally constructing more complicated photonic lattice with the  synthetic frequency dimension, showing important promise for optical communications in fiber-based or on-chip resonators \cite{chen2021highlighting},  and also highlighting potentials towards  non-Hermitian/topological \cite{peng2014parity,wimmer2015observation,lulingRevModPhys,PhysRevApplied.14.064076,guo2021experimental} and quantum photonics \cite{boutari2016large,luchaoyang2018,PhysRevLett.123.150503,joshi2020frequency} in coupled modulated ring resonator systems.

%\section*{Online content}
%Any methods, additional references, Nature Research reporting summaries, source data, extended data, supplementary information, acknowledgements, peer review information; details of
%author contributions and competing interests; and statements of
%data and code availability are available at https://doi.org/...

\section*{Methods}
\textbf{Experimental setup.}
The frequency of the laser source can be finely tuned over 30 GHz by applying an external ramp signal  to its frequency modulation input, with the central wavelength located at 1550.92 nm.  Near 50$\%$ of the laser source is sent to an  acousto-optic modulation (AOM) for frequency shift and heterodyne beating with the drop-port output \cite{wildi2020photo,tetsumoto2021optically}. In each ring, a $2\times2$ fiber coupler couples 1$\%$ of the remaining 50$\%$ laser source to the ring resonator, after which a polarization controller is used to adjust the polarization of laser circulating in the ring. To achieve a high quality factor for the resonator, a semiconductor optical amplifier (SOA) is used to compensate the loss in the ring with a maximum gain of 10 dB. A dense wavelength division multiplexing  with a central wavelength of 1550.92 nm (international telecommunication union channel 33) is used to filter the amplified emission noise from SOA. Ring A undergoes dynamic modulation by  a lithium niobate EOM with a 10 GHz bandwidth, which is driven by  an arbitrary waveform generator with 200 MHz bandwidth. A $1\times2$ fiber coupler couples 0.5$\%$ of the signal  out of the ring, which is then amplified by  an erbium-doped optical fiber amplifier (with maximum gain of 12 dB) to boost the signal-to-noise ratio before it gets detected by a fast InGaAs photodiode (850 to 1650 nm with 10 GHz bandwidth) and sent to the oscilloscope (5 Gsamples/s with 1 GHz bandwidth). For the flat band structure measurement (Figs.~\ref{fig3} and \ref{fig4}), we disconnect the  AOM path, and only connect it for the resonant mode observation (Fig.~\ref{fig5}). In addition, we also place an EOM in ring B just for calibrating the length of ring B, which is not shown in Fig.~\ref{fig2}.

The lengths of the two rings need accurate calibration for stabilize the connectivity between the resonant modes to construct the synthetic Lieb lattice. We first separately measure the FSRs of ring A and ring B by disconnecting the 70:30 fiber coupler. For each ring, we vary the modulation frequency by linearly sweeping the input frequency until the modulation sidebands fully overlap with the resonant modes. We then adjust fiber's length to make up for the required FSR difference. One can also place an optical delay in the ring to finely tuning the length.  Noting that the fiber coupler used to couple the two rings together in Fig.~\ref{fig2} keeps 70$\%$ of the light power remained in the excited ring and the left 30$\%$ coupled to the other ring no matter which ring we choose to excite, which gives the same coupling strengths for both cases.

\section*{Data availability}
Source data are provided with this paper. All other data that support the
plots within this paper and other findings of this study are available from the
corresponding authors upon reasonable request.

\section*{Acknowledgements}
We greatly thank Prof. Shanhui Fan for fruitful discussions. The research is supported by National Natural Science Foundation of China (12122407, 11974245, and 12104297), National Key R$\&$D Program of China (2017YFA0303701), Shanghai Municipal Science and Technology Major Project (2019SHZDZX01), Natural Science Foundation of Shanghai (19ZR1475700),  and China Postdoctoral Science Foundation (2020M671090). L.Y. acknowledges support from the Program for Professor of Special Appointment (Eastern Scholar) at Shanghai Institutions of Higher Learning.  X. C. also acknowledges the support from Shandong Quancheng Scholarship (00242019024).

\section*{Author contributions}
G.L. conceived the idea with L.Y. and X.C., and designed the experiment. G.L., L.W. and L.Y. developed the theoretical analysis and simulations. G.L. and R.Y. carried out the experiment with assistance from S.L. and Y.Z.  L.Y. and X.C. revised the manuscript. All authors contributed to discussion of the results and preparation of the manuscript. G.L., L.Y. and X.C. supervised the project.

\section*{Competing interests}
The authors declare no competing interests.

%\section*{References}
%
%\bibstyle{jabbrv}
\bibliography{flatband-manuscript}

\end{document}